\documentclass[11pt,a4paper,twoside]{article}
\usepackage{amssymb,amsfonts,amsmath,amsthm,euscript}
\usepackage{graphicx,subfigure}
\usepackage[left = 2.9cm, right = 2.9cm, top = 3cm, bottom = 3cm]{geometry}
\usepackage{multicol}
\usepackage{slashbox, rotating, color}
\usepackage{dsfont}
\theoremstyle{definition}

\newcommand{\newfootnote}[2]{$\mbox{#1}^{\mbox{\footnotesize #2}}$}
\newcommand{\newfootnotetext}[1]{\hspace{-10pt}$\phantom{.}^{\mbox{\footnotesize #1}}$}

\begin{document}

\title{\large  \bf Amazon Forest Fires Between 2001 and 2006 \\
\vspace{0.2cm} and Birth Weight in Porto Velho \rm \vspace{1.0cm}}

\author{\small \newfootnote{Taiane S. Prass}{a},  \newfootnote{S\'ilvia R.C. Lopes}{a}, \newfootnote{Jos\'e G. D\'orea}{b}, \newfootnote{Rejane C. Marques}{c}  and \newfootnote{Katiane Brand\~ao}{d}\vspace{0.8cm}\\
\hspace{-90pt}\newfootnotetext{a}Institute of Mathematics - UFRGS, Porto Alegre - RS - Brazil \\
\hspace{-163pt}\newfootnotetext{b}Department of Nutrition - UnB, Bras\'ilia - Brazil\\
\hspace{-81pt}\newfootnotetext{c}Anna Nery Nursing College - UFRJ, Rio de Janeiro - RJ -  Brazil\\
\hspace{-155pt}\newfootnotetext{d}Medical School - UNIR, Porto Velho - RO - Brazil\vspace{0.6cm}
}

\date{July 3, 2011}

\maketitle

\let\thefootnote\relax\footnotetext{
\begin{multicols}{2}
\noindent *Corresponding author's address:  Taiane S. Prass\\
Instituto de Matem\'atica\\
Universidade Federal do Rio Grande do Sul\\
91509-900 Porto Alegre, RS -  Brazil \\

\vspace{0.4cm}

\noindent E-mail: taianeprass@gmail.com\\
\noindent FAX: 55-51-3308-7301
\end{multicols}
}

\noindent {\bf Running head:} Birth weight and forest fires in the Amazon.\\

\noindent{\itshape The authors declare they have no competing financial interests.}

\vspace{0.5cm}

\thispagestyle{empty}

\begin{abstract}
Intentional forest fires in the Amazon pose a serious environmental problem and a threat to the delicate balance of the rain forest. Besides that, there are immediate consequences affecting biodiversity and climate changes for a vast area of South America. We collected birth weight data from a public hospital in Porto Velho (West Amazon), to assess the effects of forest-fire smoke on human reproductive outcome in an urban center. We used multiple statistical models to assess the impact of forest fires on $22,012$ live-births during the years from $2001$ to $2006$. The total number of heat spots varied during the studied period; doubling from $2001$ to $2002$ then decreasing from $2002$ to $2003$, it rose again from $2003$ to $2004$ and from $2004$ to $2005$. Altogether, the number of heat spots doubled from $2003$ to $2005$.
Because difference in birth weight between sexes was statistically significant, it was included in all statistical models. Birth weight per year or per semester showed no statistically significant difference.
Testing for the combined effect of semester within a year showed no significant difference for girls, but for boys the significant difference was attributed to analytical artifact. However, when we tested for birth weight as a function of months of a determined year there were significant differences for both boys and girls; the significant difference corresponded to months (March and November of 2003) with the highest number of heat spots.
\end{abstract}

\section{Introduction}

The aggressive occupation of the Brazilian Amazon dates back to the expansion of the agricultural frontier that started in the $1970$s with the opening of roads. Agriculture, mining, and economic development have led to human occupation and emergence of urban centers in the West Amazon; the state of Rondonia bordering Bolivia, is one of the most impacted by forest fires due to its expanding agriculture and mining activities. Government incentives boasted large-scale agriculture projects, always starting with forest clearances by fire (Arag\~ao and Shimabukuro, 2010). Agricultural land claimed from the felled forest has been made into cropland and pasture, which is managed by fire. Despite official efforts to limit further forest land occupation or preservation, the increasing price of meat (green meat) and soy beans has driven forest destruction at alarming rates in the last $10$ years (Arag\~ao and Shimabukuro, 2010); all done by using deliberate but out of control forest fires.  The extent of the felled native forest and, the continuing ecological damage to the native flora and fauna remain unknown. However, there have been attempts to assess health impacts on some urban centers.

\vspace{0.2cm}

Ignotti et al. (2010) showed an association of atmospheric pollution (due to forest burning)
with occurrences of respiratory diseases (in children and the elderly) in the Brazilian
Amazon region. Respiratory diseases in the elderly due to forest fires accounted for $50$ to $80\%$
of mortality rate in the state of Rondonia between the years of $1998$ and $2005$ (Castro et al., 2009). Also in the West Amazon, Mascarenhas et al. (2008) measured particulate matter $< 2.5$microm (resulting from forest fire) and showed its association with increased rates of respiratory diseases in children $<10$ years of age.

\vspace{0.2cm}

This study is a first attempt to assess the impact of atmospheric pollution brought about
Amazon forest fires on reproductive outcomes (birth weight) in a large urban
center - the city of Porto Velho - located in the most impacted region of deforestation
and human occupation in the Amazon.

\section{Materials and Methods}

\vspace{0.4cm}

\noindent {\bf Data acquisition \rm}

\vspace{0.4cm}

This is part of our ongoing project on environmental hazards to children's health
in the West Amazon (D\'orea et al., 2007); the research protocol was approved
by the Ethics Committee of Studies for Humans of the Universidade Federal de
Rondonia. We had access to the birth records of children born at the Hospital
de Base, the largest public hospital facility caring mostly for the lower middle
class in the city of Porto Velho (capital of the state of Rondonia, West Amazonia).
A parent publication dealing with ethylmercury exposure derived from
Thimerosal-containing vaccines given to newborns appeared elsewhere (D\'orea et al., 2009).

\vspace{0.2cm}

We collected data of live births occurring from $2001$ to $2006$; these data were recorded
by hospital staff and comprised live birth-weight, sex, and date of birth. The total
number of observations for the period was $n = 22012$ with a sex proportion of $48.4\%$
of females and $51.6\%$ males. Forest fire information was based on data collected
by the satellite NOAA-12 with records summarized monthly.

\newpage


\noindent {\bf Data summarization \rm}

\vspace{0.4cm}

\begin{figure}[htbp]
    \centering
    \includegraphics[width = 1.05 \textwidth, height = 0.20\textheight]{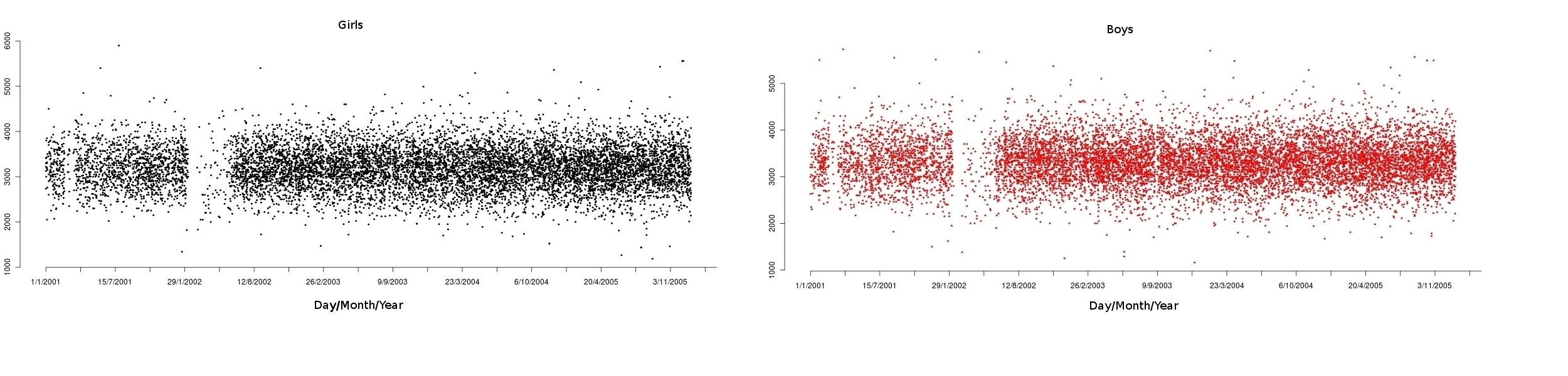}\vspace{-0.3cm}
    \caption{Birth weight for girls  (black) and boys
(red), in the period from January 2001 to December
2005.}\label{weightsex}
    \end{figure}

Figure \ref{weightsex} shows the observed values of the birth
weights for girls (left plot) and for boys (right plot), during the
studied period. The lack of observations during the period from
12/02/2002 to 02/03/2002 corresponds to interrupted services at the
maternity ward due to building repairs on the hospital.

\vspace{0.1cm}

The descriptive statistics for the birth weight data are
presented in Table \ref{tabledescriptive}. The range of variation is wider for girls
($4715$g) than for boys ($4570$g); moreover, both, minimum  and maximum
weights are higher for girls than for boys (see Table \ref{tabledescriptive}). Nevertheless,
the coefficient of variation (CV) is relatively low for both girls and boys.

\begin{table}[htb]
\centering \renewcommand{\arraystretch}{1.1} \caption{Descriptive
statistics of the birth weight by sex in the period of
2001-2005.}\label{tabledescriptive} \vspace{.2cm}
\begin{tabular}{c|cccccccc}
\hline \hline
Sex &Min &1st Q &Median &  Mean &3rd Q &   Max & DP  & CV \\
\hline
Female &   1185 &   2900  &  3200 &   3193 &   3490  &  5900 & 450.8375& 0.1412\\
Male &   1160 &   3000 &   3305  &  3310  &  3600  &  5730 & 477.6589& 0.1443\\
\hline
\hline
\end{tabular}
\end{table}

\begin{figure}[!htb]
    \centering
    \mbox{
    \subfigure[]{\includegraphics[height = 0.22\textheight]{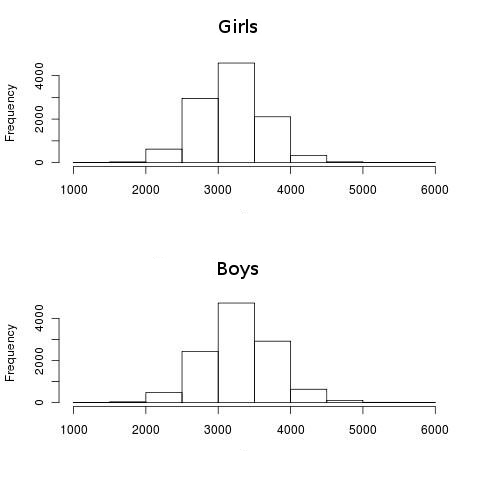}}
    \subfigure[]{\includegraphics[height = 0.22\textheight]{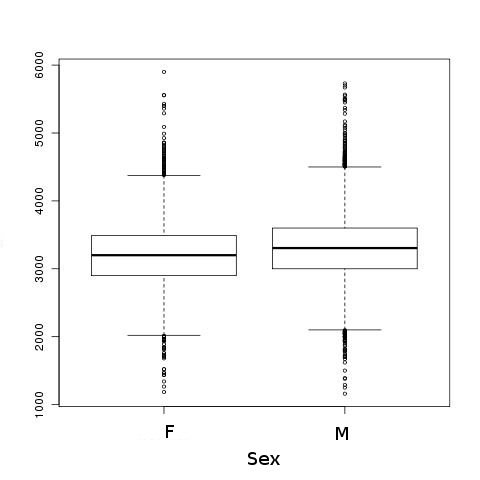}}}
    \caption{(a) Histogram and (b) box-plot  of the weights by
      sex.} \label{Histogram}
    \vspace{-2cm}
    \end{figure}

The data are also illustrated in Figures \ref{Histogram}(a) and 2(b) as a
histogram and a box-plot, respectively. While the histogram shows an
overall symmetry between girls and boys regarding ranges of birth
weight, the box-plot indicates the presence of outliers in the data.
The frequency of outliers for boys is $1.66\%$ and, for girls is $1.02\%$.

\newpage

\begin{figure}[!ht]
\centering
\includegraphics[scale = 0.35]{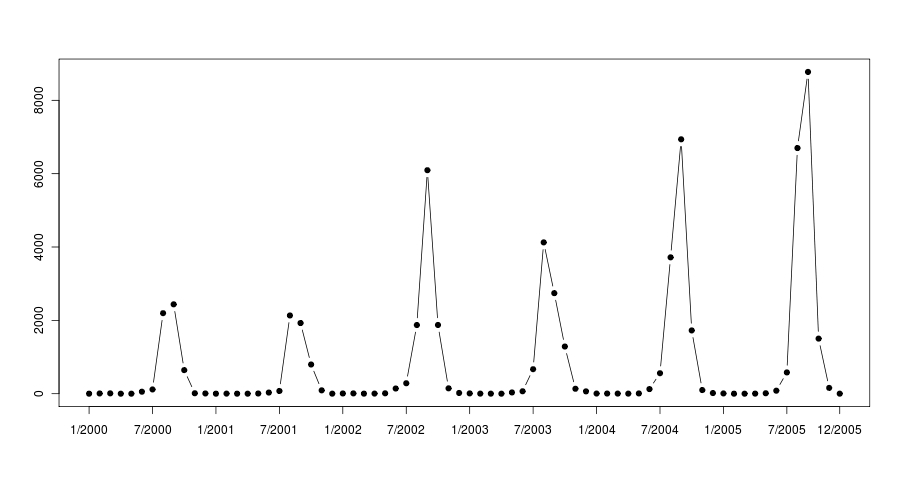}
\caption{Number of heat spots by month in the period of January 2000 to  December 2005, in Rondonia.}\label{Heatspots}
\end{figure}

Table \ref{tableheatspots} shows a summary of heat spots captured by satellite
images for the state of Rondonia during the studied period; the total number of
heat spots varied on a yearly basis, almost doubling from $2001$ to $2002$ then decreasing from $2002$ to $2003$, rising again from $2003$ to $2004$ and from $2004$ to $2005$. Altogether, the number of heat spots practically doubled from $2003$ to $2005$. An illustration of heat spots is graphically shown in Figure \ref{Heatspots}; it is worth noting that their occurrence is higher during the second semester, which is the dry season (see Figure \ref{Heatspots} and Table \ref{tableheatspots}). In tandem with this observation, if forest fires affected birth weights more light children would be expected during the first semester than in the second semester. Furthermore, for years with a higher number of heat spots one would also expect more light birth weights.

\begin{table}[!ht]
 \centering
 \renewcommand{\arraystretch}{1.1}
 \renewcommand\tabcolsep{1.5pt}
\caption{Number of heat spots captured by satellite images for the state of Rondonia during the studied period.} \label{tableheatspots}
\vspace{.2cm}

{\footnotesize\begin{tabular}{c|cccccccccccc|r}
\hline
\hline
Year & {\scriptsize January} & {\scriptsize February} & {\scriptsize March} & {\scriptsize April} & {\scriptsize May} & {\scriptsize June} & {\scriptsize July} & {\scriptsize August} & {\scriptsize September} & {\scriptsize October} & {\scriptsize November} & {\scriptsize December} & Total\\
\hline
2000 & 0 & 6 & 9 & 0 & \ 0 & \ 54 & 116 & 2196 & 2437 & \ 643 & \ 12 & \ 6 & 5485 \\
2001 & 0 & 2 & 0 & 0 & \ 5 & \ 31 & \ 74 & 2133 & 1927 & \ 796 & \ 91 & \ 1 & 5060 \\
2002 & 6 & 8 & 1 & 3 & 10 & 141 & 285 & 1872 & 6095 & 1874 & 146 & 20 & 10461 \\
2003 & 8 & 2 & 1 & 0 & 34 & \ 66 & 668 &  4126 & 2742 & 1286 & 134 & 63 & 9130 \\
2004 &4 & 4 &3 & 2 & \ 7 & 126 & 558 & 3718 & 6937 & 1725 & \ 99 & 17 & 13200 \\
2005 & 7 & 0 & 0 & 2 & 12 & \ 82 & 580 & 6699 & 8775 & 1502 & 158 & \ 1 & 17818 \\
\hline
Total & 25 & 22 & 14 & 7 & 68 & 500 & 2281 & 20744 & 28913 & 7826 & 640 & 108 & 61154 \\
\hline
\hline
\multicolumn{14}{l}{{\footnotesize Source:  INPE -  Satellite   NOAA-12}}
\end{tabular}}
\end{table}


It can be noticed that the greatest number of fires occurred during the
months of August and September of any given year, which correspond to
the peak of the dry season. Considering that the last semester of
pregnancy could be the most determinant of birth weight, it is expected
that March to June should show effect. We speculate that the smoke pollution from forest
fires can be unfavorable for the first months of a pregnancy because they could affect the fetus
initial development, thus modulating weight gain throughout pregnancy.
Besides, a quite well developing fetus has more metabolic chances of gaining weight
than one with an initial developmental hindrance.

\vspace{0.2cm}

In order to identify a possible relation between the numbers of heat
spots that occurred during pregnancy and related birth weight we
considered a multi-step approach. First we used ANOVA to identify
whether the sex or the semester, the month or the year of birth,
had any influence on the birth weight. After that we considered a
regression model that took into account the information obtained in
the first step to explain the mean birth weights as a
function of the number of heat spots.

\section{Results and Discussion}

As a result of the apparent differences shown in Table
\ref{tabledescriptive}, a preliminary analysis to assess the
influence of sex was necessary in order to establish if further
analysis would have to consider this variable in the statistical model.
We tested this hypothesis and, indeed, there was a statistically
significant difference between the means (F = 349.19 with p-value
$<$2.2e-16 at a confidence level of $99.9\%$). Due to sex effects on
birth weight, we included it in the analysis to assess the
influence of the heat spots on birth weight. Due to the variation
in number of heat spots over the years, several ANOVA tests were
performed in order to identify the time (year) effect on birth
weight; each sex was analyzed independently.

\vspace{.2cm}

When considering year as the only variable we cannot reject the null hypothesis that annual means are equal in both cases (girls: F = 0.4854,
p-value = 0.7465; boys: F = 1.0118, p-value = 0.3997). Because heat spots occur during the dry season, the time effect was also analyzed as a function of semesters only. In this case also no statistically significant differences between the means for the first and second semesters (girls: F = 1.814, p-value = 0.1781; boys: F = 2.3279, p-value = 0.1271).

\vspace{0.2cm}

\begin{figure}[!htb]
      \vspace{-0.2cm}
\centering
\mbox{
\subfigure[]{\includegraphics[height = 0.25\textheight]{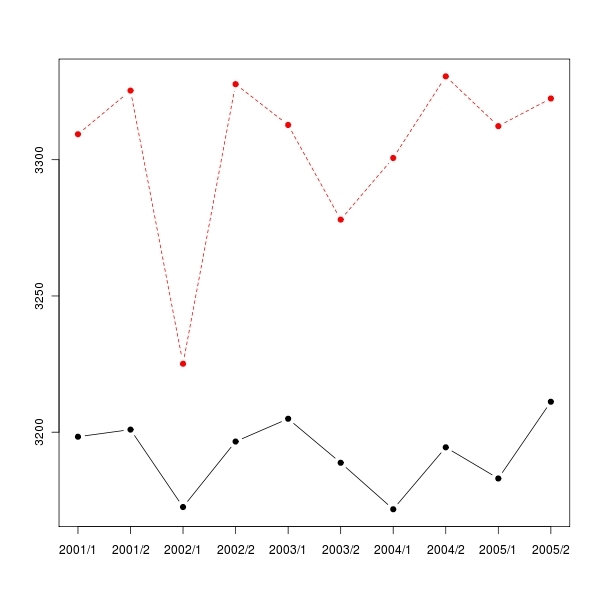}}
\subfigure[]{\includegraphics[height = 0.25\textheight]{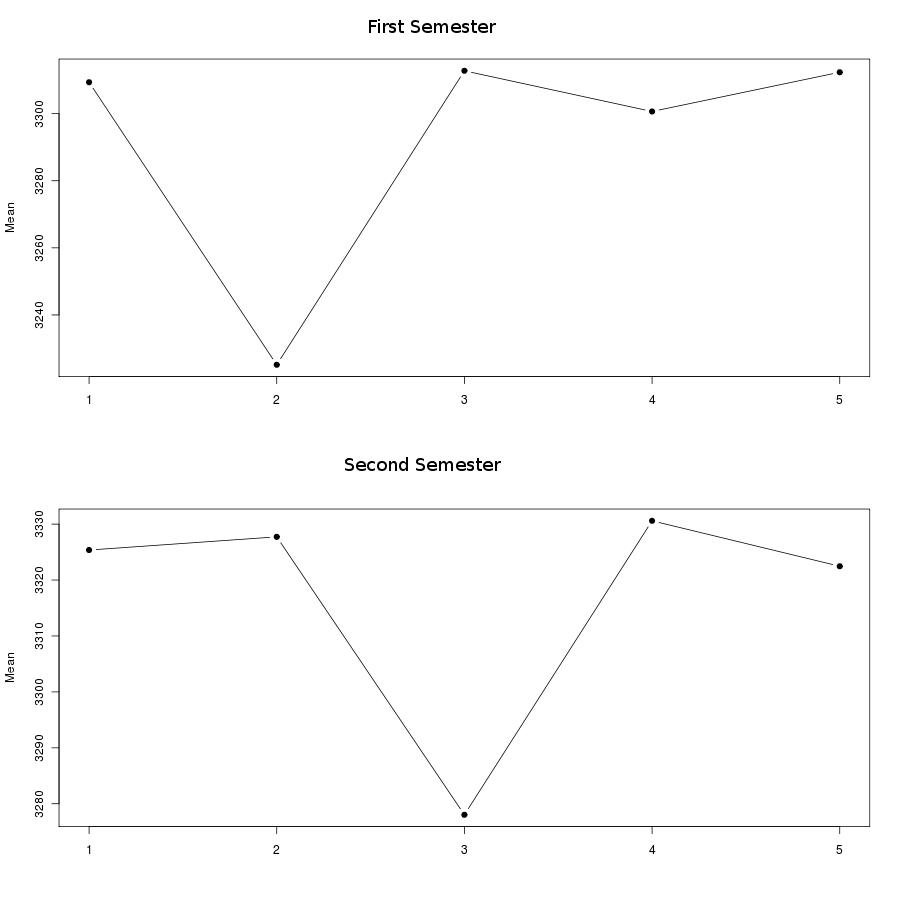}}
}
    \caption{(a) Mean weight by  semester:  in red, for boys and, in black,  for
      girls;  (b) Mean weight  by semester for boys: First semester (upper plot)
      and  second semester (lower plot). }\label{meanweight}
    \end{figure}

Figure \ref{meanweight}(a) shows mean birth weight per semester for
each studied year. One observes that only in 2003  the average in the first semester is higher than the average in the second semester for both sexes. Figure \ref{meanweight}(b) shows the mean birth weight by
year for each fixed semester. It is worth noting that for the boys
there was a noticeable decrease in birth weight in the first
semester of $2002$ that might be a bias introduced due to lower number
of observations.

The analysis done independently for semesters and
for years showed no statistically significant difference; therefore,
we decided to test the combined effects of year and semester in a
single statistical model. In this test we hoped to scrutinize for
relationship of birth weights and number of heat spots occurring in a
specific period of a specific year. In that way if there was a mean
birth weight of a semester that would differ from another, it could
possibly be detected.

    For girls, when we consider the semester by year there was no statistically
    significant effect (F = 0.8683, p-value = 0.553); also when we analyze a
    fixed semester and compare the mean for different years, there was also no
    statistically significant difference (F = 1.0743, p-value = 0.3674) for the
    first and second semesters (F = 0.4282, p-value = 0.7884).  However, for
    boys, using a similar model, there was a statistically significant effect (F
    = 2.7001, p-value = 0.0039). As shown in Figure \ref{meanweight}(a), the
    mean differing from the others is the one from the first semester of
    $2002$. Because of the confounding of fewer observations in that specific
    semester (first semester of $2002$) these results are considered an
    analytical artifact. By comparing the means for different years in a fixed
    semester, we can reject the null hypothesis that they are equal (F = 2.7583
    with p-value = 0.0263 and for the second semester, F = 2.72554 with p-value
    = 0.0278); Figure \ref{meanweight}(b) clearly shows an anomaly in the first
    semester of $2002$, and in the second semester of $2003$. Although we
    realize that the low mean weight observed during the first semester is
    unlikely to be related to forest fires the mean birth weight observed in the
    second semester of $2003$ could possibly be related to the high number of
    heat spots ($34$) registered in the first semester of $2003$.

\vspace{0.2cm}

\begin{figure}[!htb]
    \centering
    \includegraphics[width = .8\textwidth, height = 0.3\textheight]{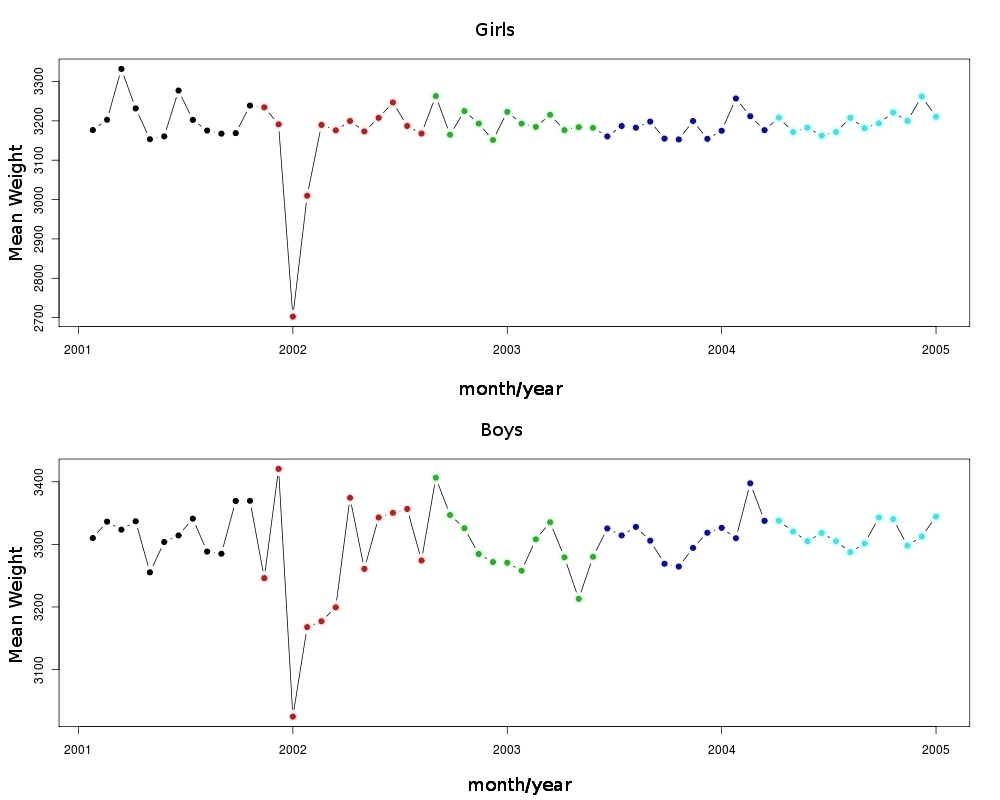}
    \caption{Mean weight by month and year:   Girls (upper); Boys  (lower).}\label{meanweightmy}
    \end{figure}

\begin{figure}[htb]
    \centering
    \includegraphics[width = .8\textwidth, height = 0.40\textheight]{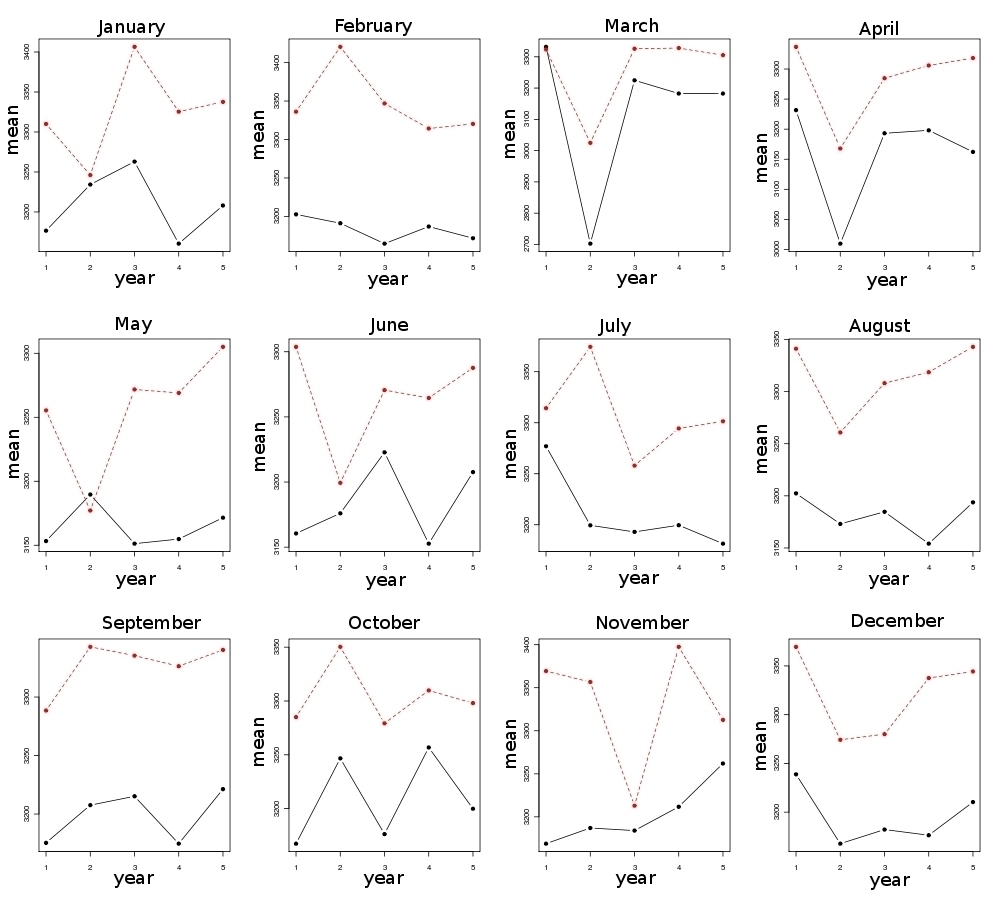}
    \caption{Mean weight by  month  for each year for  boys (dotted line, in red) and  girls.}\label{monthmeansexes}
    \end{figure}

Figures \ref{meanweightmy} and \ref{monthmeansexes} illustrate mean
birth weight per month by year; the effects of months throughout the
studied years showed a significant effect for both sexes (girls: F =
1.2440, p-value = 0.0989; boys: F = 1.5104, p-value = 0.0079). Figure \ref{meanweightmy} shows
that for both girls and boys, the salient mean birth weight
corresponds to March-2002. Considering months as a fixed effect over
the years there was no significant difference between the means, except
for March, in the case of girls (F = 5.9700, p-value = 9.938e-05) and of
boys in March (F = 2.648, p-value = 0.0323) and November (F = 4.8289, p-value =
0.0007). It should be noted that the semester mean (as a function of
year) showed significant differences; indeed, from Figure \ref{monthmeansexes}, one
observes that for March, this was the case with both boys and girls. From Table
\ref{tableheatspots} it can be seen that March-$2003$ showed the highest
number of heat spots, compared to March from other years.

Mean birth weight was statistically different between boys and girls; therefore we used a model for each sex to explain mean birth weight, not individual birth weight. The models used for girls and boys accounted for month, semester, and year of birth as well as number of heat spots. For the variable `number of heat spots' we considered number of cases corresponding to a gestation period and overall number for a specified gestation period. After a careful step-by-step run of several statistical models we could conclude that:

\vspace{0.2cm}

\noindent {\bf Girls:}
\begin{itemize}
\item[{\rm a)}] Mean birth weight for girls does not depend on month, semester, or year when these variables are analyzed separately.

\item[{\rm b)}] When mean birth weight is analyzed as a function of a specific month and year, a significant difference is detected for March-2002. However, further  detailed analysis showed that introducing a constant for March and April of 2002 explains variations in birth weight for girls. As a consequence there was no statistically significant evidence that birth weight for girls can be affected by number of forest fires.
\end{itemize}

\vspace{0.2cm}

\noindent {\bf Boys:}
\begin{itemize}
\item[{\rm a)}] Overall, when the variables month, semester, and year are considered per se, there is no statistically significant difference. However, further detailed analysis showed a significant factor influencing mean birth weight for the period
    from March to June of 2002.
\item[{\rm b)}] In order to introduce the `number of heat spots' variable in the model, we included it
as a new variable, denoted by $N(\cdot)$, and tested the following possibilities:
    \begin{itemize}
      \item $N(\cdot)$ = Number of heat spots in $i$-th month of gestation, where {\small $i \in \{0,1, \cdots, 9\}$},
    \item $N(\cdot)$ = Sum of the number of heat spots from month $i$ to $j$, with $i \leq j$  and {\small $i, j \in \{0,1, \cdots, 9\}$},
    \end{itemize}
    where, by definition,   `0'  is  the month at which  pregnancy  has started.
\end{itemize}

\vspace{0.2cm}

\begin{figure}[!ht]
\centering
\mbox{
\subfigure[Girls]{\includegraphics[width = 0.45\textwidth, height = 0.15\textheight]{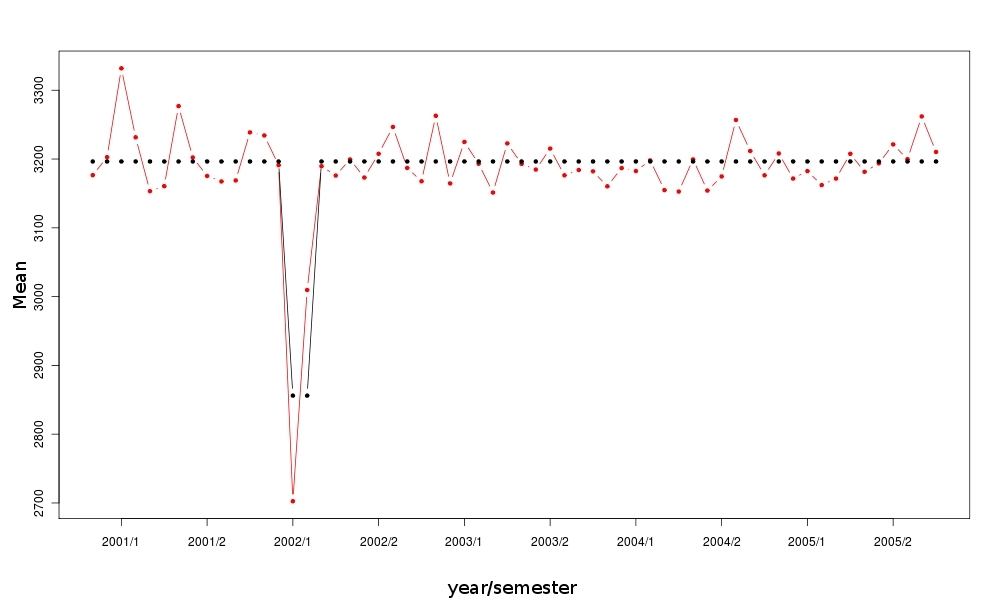} }
\subfigure[Boys]{\includegraphics[height = 0.15\textheight, width = 0.45\textwidth]{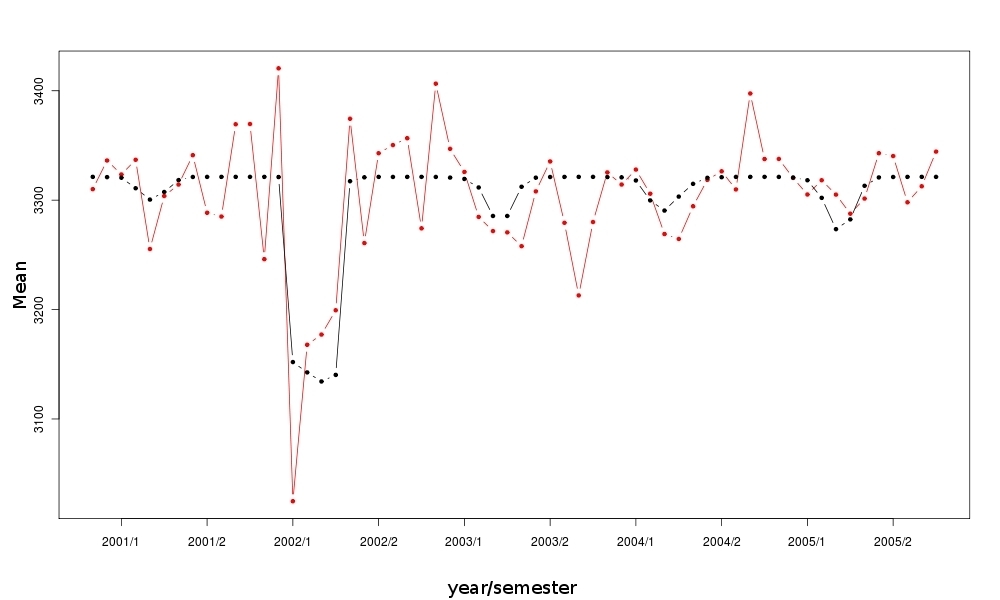}}
}
\caption{Observed monthly mean birth weight for girls and boys (red) and estimated values  (black).}\label{figmodel}
\end{figure}

Among all tested models, the best to explain mean birth weight variation (considering the value of the statistic $R^2$ in the regression model) were:

\vspace{0.2cm}

\begin{itemize}
\item {\bf For girls:} $W(m) = \mu + a\,\mathds{I}_{A}(m) = 3196.454  -340.404\,\mathds{I}_{A}(m)$,
\noindent
  where $m$ is the birth period,  $A = \{\mbox{March}-2002, \mbox{April}-2002\}$ and $\mathds{I}_{A}(\cdot)$  is the indicator function of set $A$, which is defined by $\mathds{I}_{A}(m) =1$,  if  $m\in A$,  and  $\mathds{I}_{A}(m) =0$,  if $m\notin A$.  The coefficients of the model, its p-value, and the $R^2$ value  are shown in Table \ref{girlsmodel}. The value of $R^2$ indicates that $65\%$ of the variation of the mean birth weight for girls can be explained by this model. Figure \ref{figmodel}(a) illustrates the girls' monthly mean birth weight (in red) and respective values estimated by the model (in black) for the period of $2002$ to $2005$.

\item {\bf For boys:} $W(m) = \mu + a\,\mathds{I}_{A}(m)  + bN(m) = 3321 -168.8\,\mathds{I}_{A}(m)  -0.004485N(m)$,
\noindent
 where $m$ is the birth period, $A = \{\mbox{March}-2002, \mbox{April}-2002, \mbox{May}-2002, \mbox{June}-2002\}$,  $\mathds{I}_{A}(\cdot)$ is the indicator function of set $A$, and the $N$ variable is defined as $N(m) = \sum_{ j= 0}^1 n_j$, where $n_j$ is the number of heat spots after the $j$-th month of gestation and $n_0$ is the gestation initial  month. The coefficients of the model, its p-value, and the $R^2$ value are shown in Table \ref{boysmodel}. The value of $R^2$ indicates that $54\%$ of the variation of the mean birth weight for boys can be explained by this model. Figure \ref{figmodel}(b) illustrates the boys' monthly mean birth weight (in red) and respective values estimated by the model (in black) for the period of $2002$ to $2005$.
\end{itemize}

The models captured the most accentuated decreases of mean birth weight
(Figures \ref{figmodel}(a) and \ref{figmodel}(b)) however it failed to capture the
birth weight fall of November-$2003$.

\begin{table}[!ht]
\caption{Final model for Girls mean birth weight}\label{girlsmodel}\vspace{0.2cm}
\centering
\begin{tabular}{c|crc|c}
\hline\hline
{Coefficients} & Estimate & Std. Error & t value & $\mbox{Pr($> |t|$)}$\\
\hline
Intercept &3196.454 &     5.896  &542.13 & $<$ 2e-16 ***\\
$a$       &     -340.404    & 32.294 & -10.54 &4.25e-15 ***\\
\hline
\multicolumn{5}{l}{Residual standard error: 44.9 on 58 degrees of freedom (df)}\\
\multicolumn{5}{l}{Multiple R-squared: 0.657;   Adjusted R-squared: 0.6511 }\\
\multicolumn{5}{l}{F-statistic: 111.1 on 1 and 58 df,  p-value = 4.254e-15 }\\
\hline\hline
\multicolumn{5}{l}{\footnotesize {\bf Note:} Significance code:  0 `***' }
\end{tabular}
\end{table}

\begin{table}[!ht]
\centering
\caption{Final model for Boys mean birth weight}\label{boysmodel}\vspace{0.2cm}
\begin{tabular}{c|rrr|c}
\hline\hline
{Coefficients}& Estimate & Std. Error & t value & $\mbox{Pr($> |t|$)}$\\
\hline
Intercept &  3.321e+03  &6.356e+00 &522.566 & $<$ 2e-16 ***\\
$a$       &    -1.688e+02  &2.178e+01  &-7.749 &1.79e-10 ***\\
$b$       &     -4.485e\,-\,03 & 2.168e\,-\,03 & -2.069 &  0.0431 *  \\
\hline
\multicolumn{5}{l}{Residual standard error: 41.92 on 57 degrees of freedom (df)}\\
\multicolumn{5}{l}{Multiple R-squared: 0.5429;  Adjusted R-squared: 0.5269 }\\
\multicolumn{5}{l}{F-statistic: 33.85 on 2 and 57 df,  p-value = 2.044e-10 }\\
\hline\hline
\multicolumn{5}{l}{\footnotesize {\bf Note:} Significance codes:  0 `***' 0.001 `**' 0.01 `*' }\end{tabular}
\end{table}

\vspace{0.4cm}



The smoke and particulate matter generated by forest fires currently
going on in the Amazon have been positively associated with morbidity
and mortality of vulnerable groups (Mascarenhas et al., 2008; Castro et al., 2009;
Ignotti et al., 2010). These effects are caused by impairment of lung function
due to physicochemical properties of the inhaled particles (Schwarze et al., 2006).
Population based studies have shown an increased risk of low birth weight associated
with households that use biomass fuel (Boy et al., 2002; Tielsch et al., 2009;
Thompson et al., 2011).

\vspace{0.2cm}

We noticed that, all analysis indicating that the means are not statistically equal,
also indicated that differences occurred in the means corresponding to the period with missing observations.
Dealing with missing data is always challenging and, based on these findings, we believe that a more
complete data set could lead to different models and attendant outcomes. For instance, the indicator function, included in both
models to account for the information that is not available, would not be necessary.

\vspace{0.2cm}

We could not control for variables known to interact with reproductive outcomes.
On a population based birth weight outcome there was no direct association of a
particular year or season. However, our study showed air pollution (derived from
forest fires) to have an unfavorable association with birth outcomes.

\section*{Acknowledgments}

Taiane S. Prass was partially supported by CNPq-Brazil.
S.R.C. Lopes was partially supported by CNPq-Brazil, by CAPES-Brazil,
by INCT {\it em Matem\'atica} and by Pronex {\it Probabilidade e
Processos Estoc\'asticos} - E-26/170.008/2008 -APQ1.

\end{document}